\documentclass[aps,prb,reprint,superscriptaddress,showpacs]{revtex4-1}
\usepackage{graphicx}

\usepackage{dcolumn}% Align table columns on decimal point
\usepackage{bm}% bold math
\usepackage{amsmath, amssymb, amsfonts}
\usepackage{multirow}
\usepackage{tabularx}
\usepackage[sort&compress]{natbib}
\setcitestyle{super}
\begin{document}

%\preprint{AIP/123-QED}

%\title[Determination of the indirect band gaps of rhombohedral and tetragonal thin film BiFeO$_3$ phases]{Determination of the indirect band gaps of rhombohedral and tetragonal thin film BiFeO$_3$ phases}% Force line breaks with \\
\title[Anisotropic optical properties of rhombohedral and tetragonal thin film BiFeO$_3$ phases]{Anisotropic optical properties of rhombohedral and tetragonal thin film BiFeO$_3$ phases}%
%\affiliation{Singapore Synchrotron Light Source, National University of Singapore, 5 Research Link, Singapore 117603, Singapore}
%\affiliation{NUSNNI-Nanocore, Department of Physics, National University of Singapore, 2 Science Drive, Singapore 117551, Singapore}
%\affiliation{School of Materials Science and Engineering, Nanyang Technological University, Singapore 639798, Singapore}

%tetragonal -> cubic base 'a' and 'a' not equal to 'c' (out-of-plane); BFO on LAO
%rhombohedral -> a = b = c; alpha = beta = gamma != 90; BFO on STO

\author{Daniel Schmidt}
 \email{schmidt@huskers.unl.edu}
 \affiliation{Singapore Synchrotron Light Source, National University of Singapore, 5 Research Link, Singapore 117603, Singapore}
\author{Lu You}
 \affiliation{School of Materials Science and Engineering, Nanyang Technological University, Singapore 639798, Singapore}
\author{Xiao Chi}
 \affiliation{Singapore Synchrotron Light Source, National University of Singapore, 5 Research Link, Singapore 117603, Singapore}
%\author{Pranjal K. Gogoi}
% \affiliation{Singapore Synchrotron Light Source, National University of Singapore, 5 Research Link, Singapore 117603, Singapore}
% \affiliation{NUSNNI-Nanocore, Department of Physics, National University of Singapore, 2 Science Drive, Singapore 117551, Singapore}
\author{Junling Wang}
 \affiliation{School of Materials Science and Engineering, Nanyang Technological University, Singapore 639798, Singapore}
\author{Andrivo Rusydi}
 \email{phyandri@nus.edu.sg}
 \affiliation{Singapore Synchrotron Light Source, National University of Singapore, 5 Research Link, Singapore 117603, Singapore}
 \affiliation{NUSNNI-Nanocore, Department of Physics, National University of Singapore, 2 Science Drive, Singapore 117551, Singapore}

\date{\today}

\begin{abstract}
The anisotropic optical properties of multiferroic BiFeO$_3$ thin films have been determined with Mueller matrix ellipsometry at room temperature. The full dielectric tensors of tetragonal-like and rhombohedral-like BiFeO$_3$ phases epitaxially grown on LaAlO$_3$ and SrTiO$_3$ single crystal substrates, respectively, within the spectral range of 0.6 and 6.5~eV are reported. Strain-driven anisotropy changes and transition shifts are observed as well as evidence of sub-band gap many-particle excitations are found. The transition shifts, mostly to higher energies for the highly strained tetragonal-like BiFeO$_3$ phase on LaAlO$_3$, are indicative of band structure differences. Additionally, optical modeling, confirmed by piezoelectric force microscopy studies, revealed that the average polarization direction of bivariant BiFeO$_3$ on LaAlO$_3$ is not parallel to the crystallographic [001] direction but tilted by about $7^{\circ}$. Spectral weight analyses reveal phase-dependent differences, underlining that theoretical calculations of optical spectra need further improvement to appropriately account for electronic and excitonic correlations to fully understand multiferroic BiFeO$_3$.
\end{abstract}
% The indirect and direct band gaps are determined and strain-driven band structure changes are observed.


\maketitle

%Spectral weight analyses show that electron correlations are phase-dependent and ab-initio density functional theory calculations need further improvement to fully understand multiferroic BiFeO$_3$.

%many body effects
%neutral excitations
%calculation of optical properties/spectra

\section{Introduction}
Single crystalline bismuth ferrite (BiFeO$_3$) is a multiferroic perovskite structure and exhibits magnetic as well as strong ferroelectric behavior at room temperature. For about the past decade, BiFeO$_3$ has been of strong research interest due to its potential applicability in ferroelectric memory devices and spintronics as well as photovoltaics, for example~\cite{Catalan2009}.
While the lattice system of bulk BiFeO$_3$ is rhombohedral, the crystal structure of thin films can be engineered by introducing epitaxial strain. Depending on the choice of single crystalline substrate materials and their different lattice parameters, the thin film BiFeO$_3$ crystal structure and associated physical properties, such as transition energies, can be modified~\cite{Wang2003,Chen2010NTU,Chen2011}.

Although there are already numerous publications about the optical properties of bulk or thin film BiFeO$_3$, often the highly anisotropic nature of crystalline BiFeO$_3$ has not been appropriately considered during experiment and data analysis. Apart from that, especially the optical band gap energy and its absorption onset, and whether BiFeO$_3$ is a direct or indirect material is controversially discussed.

Most \emph{ab initio} calculations agree that BiFeO$_3$ is a semiconductor with an indirect band gap that is very close to the first direct transition due to the flatness of the bands~\cite{Neaton2005}. However, while some argue that the closeness of direct and indirect gaps is due to the valence band being very flat~\cite{Clark2007,Palai2008}, others have calculated very flat conduction bands~\cite{Wang2009,Liu2011b}.
In general, \emph{ab initio} calculations to characterize multiferroics with transition metal cations and oxygen may be very complicated as both the exchange interaction and electron correlations have to be taken into account. Additionally, excitons are typically not considered when calculating optical properties, even though excitonic effects may strongly influence the dielectric function tensor~\cite{Gogoi2015}. Given the multitude of available variations of computational modeling with density functional theory (DFT), different results are not surprising~\cite{Stroppa2010}.

Two experimental reports regarding polycrystalline BiFeO$_3$ thin films suggest the presence of an indirect gap roughly 1.0 eV below the first direct transition~\cite{Gujar2007,Fruth2007}. However, many others argued that no indications of an indirect gap were found, and concluded that BiFeO$_3$ is a direct band gap material with a transition energy between 2.6 and 3.1~eV at room temperature (mainly depending on the crystalline phase)~\cite{Basu2008,Hauser2008,Ihlefeld2008,Kumar2008,Himcinschi2010,Chen2010,Choi2011,Liu2013,Himcinschi2015}.

Interestingly, all of these aforementioned references point out that the absorption onset starts significantly below the first direct transition. One of the reasons for this can be a shallow oxygen vacancy state below the conduction band, as calculated by Clark and Robertson~\cite{Clark2009} and as is consistent with observations by Hauser \emph{et al.}~\cite{Hauser2008}. Ju and Cai showed in a theoretical study that the absorption onset might strongly redshift with increasing defect states in the form of oxygen vacancies~\cite{Ju2009b}. However, even though a redshift of the first allowed direct transition was confirmed with increasing oxygen vacancies a shift of the absorption onset could not be observed experimentally~\cite{Jiang2011}.
The early absorption onset far below the main transition (often observed as a long structureless enhanced spectral weight) was assigned by Pisarev \emph{et al.} to charge transfer instabilities accompanied by a self-trapping of excitons~\cite{Pisarev2009}. Xu \emph{et al.}, however, observed two distinct peaks at around 1.4 and 1.9~eV in transmittance measurements of a bulklike single crystal and assigned them to on-site crystal-field transitions~\cite{Xu2009}. These peaks are consistent with previously reported many-particle transition bands also involving excitons~\cite{Galuza1998}.

It is noteworthy that in most previous studies the optical properties have been determined by making use of parametrized oscillator models to extract the dielectric function from spectroscopic ellipsometry or transmittance data, for example~\cite{Kumar2008,Ihlefeld2008,Himcinschi2010,Chen2010,Choi2011,Liu2013,Himcinschi2015, Jiang2011,Pisarev2009}. Such sub-band gap transitions have not been reported in these studies. In general, with the use of parametric physical line-shape models, a certain risk is involved for subtle spectral features to be neglected by the line shape of the model function. Since the characteristics of indirect transitions are often slight and the absorption due to crystal-field transitions is approximately three orders of magnitude smaller than the absorption above the charge gap, particular care must be taken here~\cite{Xu2009}.

%This is probably the reason why none of these studies on rhombohedral-like BiFeO$_3$ has been observing such sub-band gap many-particle transitions.

Here, we present the anisotropic optical properties of multiferroic BiFeO$_3$ thin films with a nominal thickness of around 35~nm as determined with Mueller matrix ellipsometry at room temperature. Ellipsometry within the Mueller-Stokes formalism has been shown to be an excellent technique for the determination of the dielectric function tensor of biaxially anisotropic materials~\cite{Schubert1996,Schubertbook,Schmidt2013ellinanoscale,Schmidt2009a,Schmidt2009b,Makinistian2010}. The full dielectric tensors as well as major polarizability directions of epitaxial rhombohedral-like and tetragonal-like BiFeO$_3$ phases are discussed. Charge transfer transitions of both films are quantified and strain-induced differences in excitation energies between rhombohedral and highly strained tetragonal phases are reported. A comparison of experimental data from both crystalline phases with recent DFT with the Heyd-Scuseria-Ernzerhof hybrid functional is made to reveal that many-body effects are not yet appropriately accounted for and particularly excitations of an excitonic nature need to be taken into account when calculating the dielectric function tensor~\cite{Dong2013}.

%\begin{figure}[htbp]
%	\centering
%		\includegraphics[width=.5\textwidth]{spectrum_all_zoom.eps}
%		\caption{Experimental (dotted lines) and best-match calculated (solid lines) normalized Mueller matrix spectra $M_{ij}/M_{11}$ for BiFeO$_3$ on LaAlO$_3$ (top) and BiFeO$_3$ on SrTiO$_3$ (bottom). The data is plotted for an angle of incidence $\Phi_{\rm a}=70^{\circ}$ and in-plane orientations of $\varphi=20^{\circ}$ and $\varphi=40^{\circ}$ for the LaAlO$_3$ and SrTiO$_3$ samples, respectively. Note the two axis breaks to emphasize the off-diagonal Mueller matrix elements.}
%	\label{fig:spectrum}
%\end{figure}

\begin{figure}[tbp]
	\centering
		\includegraphics[width=.4\textwidth]{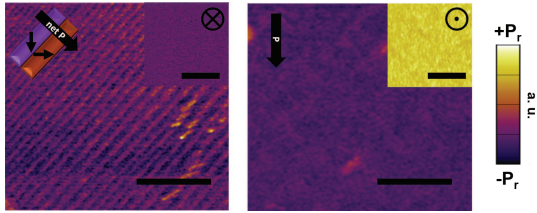}
		\caption{In-plane PFM images of BiFeO$_3$ on LaAlO$_3$ (left) and SrTiO$_3$ (right). The schematic insets in the top left corner depict the polarization directions. The insets in the top right corner show the respective single-domain out-of-plane images with indicated polarization directions. The scale bars are 500~nm.}
	\label{fig:PFM}
\end{figure}

%\section{Theory}
\section{Sample Details}

Epitaxial BiFeO$_3$ thin films were deposited on LaAlO$_3$ (001) and SrTiO$_3$ (001) single crystal substrates by pulsed laser deposition. While the LaAlO$_3$ substrate has no intentional miscut, SrTiO$_3$ has an intentional $4^{\circ}$ miscut along the [110]$_{\rm c}$ direction. A stoichiometric target was ablated by using a KrF excimer laser ($\lambda=248$~nm) with an energy density of $\thicksim1.5$~J/cm$^2$ and a repetition rate of 10~Hz. The deposition temperature was $650~^{\circ}$C and the oxygen partial pressure was 100 mTorr~\cite{You2009}.

Figure~\ref{fig:PFM} depicts in-plane and out-of-plane (insets) piezoelectric force microscopy (PFM) images of BiFeO$_3$ films on LaAlO$_3$ and SrTiO$_3$, respectively. The left image (BiFeO$_3$ on LaAlO$_3$) shows part of one bivariant in-plane domain and the diagonal ``stripes'' are parallel to atomically flat LaAlO$_3$ terraces. Due to the lattice constant of LaAlO$_3$ ($a=3.79$~\AA), the coherent BiFeO$_3$ film is in a highly strained tetragonal-like monoclinic phase (\emph{T} phase) with a giant $c/a$ ratio (1.23)~\cite{Bea2009,Luo2013,You2014}. Based on the terrace widths of approximately 76~nm an unintentional miscut angle of $0.28^{\circ}$ along the [110]$_{\rm pc}$ direction can be calculated.
The BiFeO$_3$ film on the SrTiO$_3$ substrate is not as strained and is in a rhombohedral-like monoclinic phase (\emph{R} phase) with a $c/a$ ratio of 1.03~\cite{Kim2008}. The thin film is partially relaxed due to the large substrate miscut angle. A near perfect single-domain character with approximately 5\% other random domain variants is observed in large scale in-plane PFM images.
For both rhombohedral and tetragonal BiFeO$_3$ samples, a single-domain out-of-plane character is confirmed.

The monoclinic distortion in both cases is very small ($<2^{\circ}$) and therefore an orthorhombic lattice has been assumed for the following optical analysis~\cite{Wang2003,Chen2010NTU}.

\section{Spectroscopic Ellipsometry}

Spectroscopic Mueller matrix ellipsometry spectra within the spectral range from 0.6 to 6.5~eV in steps of 20 meV were acquired using a commercial rotating analyzer instrument with a compensator (VASE, J. A. Woollam). The samples were mounted on a precision rotation stage (RS40, Newport) to perform azimuth-dependent measurements, and the in-plane rotation angle $\phi$ was varied from $0^{\circ}$ to $320^{\circ}$ in steps of $40^{\circ}$. At each in-plane orientation, data were taken at three angles of incidence $\Phi_a$ ($50^{\circ},60^{\circ},70^{\circ}$). Such an angle-resolved measurement scheme is necessary for a complete characterization of arbitrary optically anisotropic samples~\cite{Schubert1996,Schubertbook,Schmidt2013ellinanoscale,Schmidt2009a,Schmidt2009b}.
Ellipsometric spectra ($\Psi$ and $\Delta$) for the pristine isotropic substrates were measured in the same energy range at a single in-plane orientation $\phi$~\cite{HOE,Fujiwarabook}.

The complex dielectric function for the isotropic SrTiO$_3$ substrate has been calculated by wavelength-by-wavelength inversion of the experimental data (pseudodielectric function)~\cite{Fujiwarabook}. For LaAlO$_3$ the pseudodielectric function has been further parameterized by a physical line-shape model in order to avoid experimental data noise from becoming part of the calculated dielectric function and thus any further analysis of the BiFeO$_3$ thin film~\cite{HOE,Jellison1993}.

\begin{figure}[tbp]
	\centering
		\includegraphics[width=.5\textwidth, clip, trim=10 10 40 52]{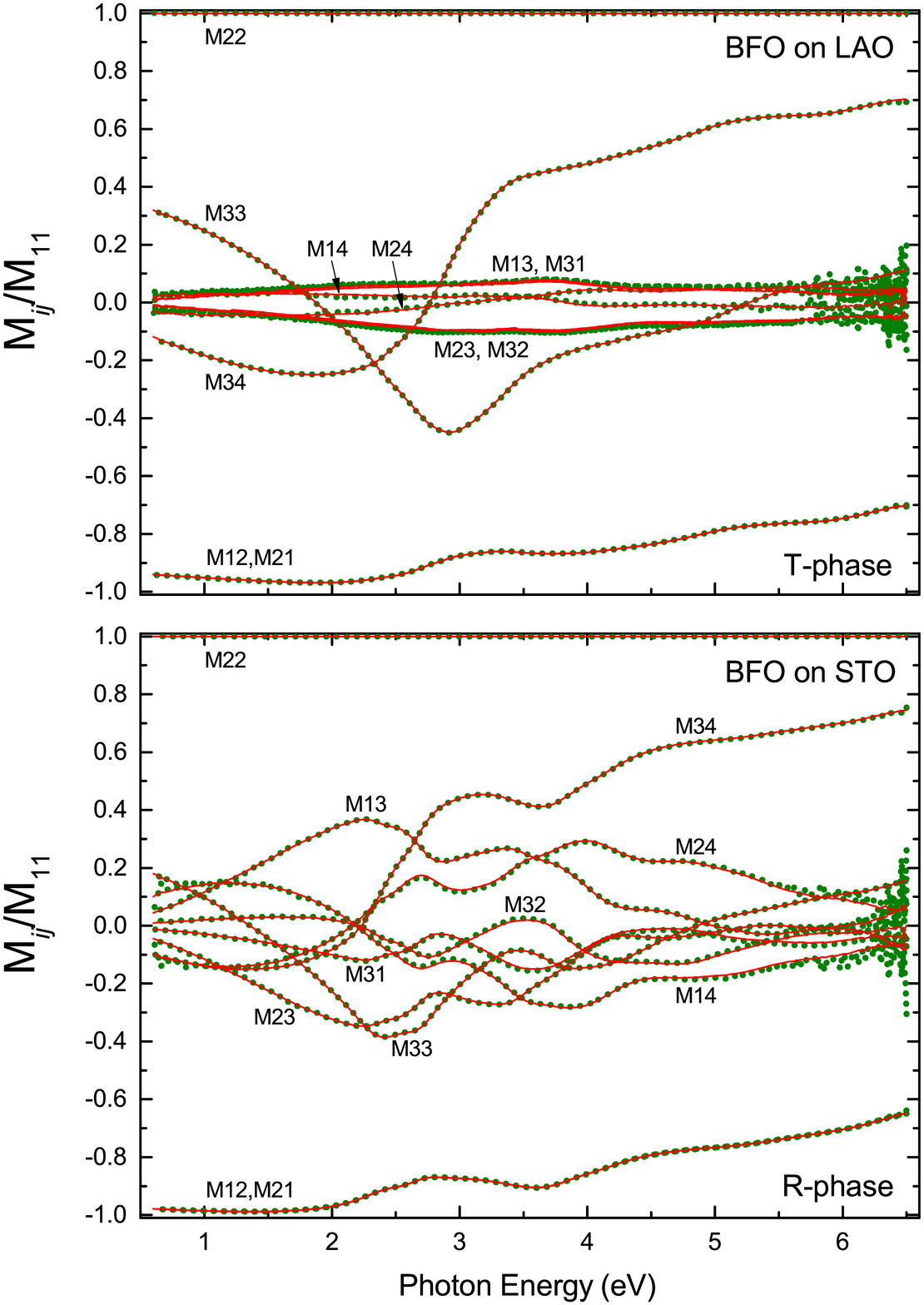}
		\caption{Experimental (dotted lines) and best-match calculated (solid lines) normalized Mueller matrix spectra $M_{ij}/M_{11}$ for BiFeO$_3$ on LaAlO$_3$ (top) and BiFeO$_3$ on SrTiO$_3$ (bottom). The data are plotted for an angle of incidence $\Phi_{\rm a}=70^{\circ}$ and in-plane orientations of $\varphi=20^{\circ}$ and $40^{\circ}$ for the LaAlO$_3$ and SrTiO$_3$ samples, respectively. Note that the off-diagonal Mueller matrix elements $M_{13}$, $M_{14}$, $M_{23}$, $M_{24}$, $M_{31}$, and $M_{32}$ are plotted $\times10$ for clarity.}
	\label{fig:spectrum}
\end{figure}

\begin{figure}[tbp]
	\includegraphics[width=.45\textwidth, clip, trim=0 0 0 0]{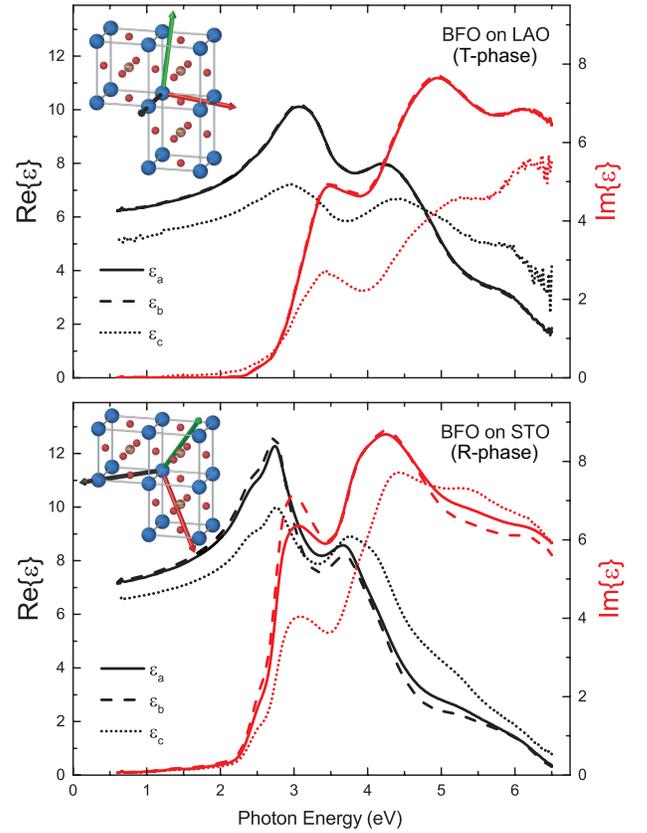}
	\caption{Real and imaginary parts of the dielectric function tensor obtained from best-match model calculations for BiFeO$_3$ on LaAlO$_3$ (top) and BiFeO$_3$ on SrTiO$_3$ (bottom). The insets show schematically the orientation of the major polarizability axes $a$ (black), $b$ (red), and $c$ (green).}
	\label{fig:eps}
\end{figure}

The stratified layer optical model for both samples under investigation comprises, besides the substrate, two additional layers accounting for the BiFeO$_3$ thin film and the surface roughness, respectively. The optical model equivalent of surface roughness is a layer with thickness $d$ and optical constants typically calculated based on a Bruggeman effective medium approximation with 50\% material and 50\% void (${\rm Re}\{\varepsilon\}=1$ and ${\rm Im}\{\varepsilon\}=0$)~\cite{Fujiwarabook}.
The BiFeO$_3$ thin film in both cases is assumed to have an orthorhombic crystal system with complex dielectric function parameters $\varepsilon_{a}$, $\varepsilon_{b}$, and $\varepsilon_{c}$ along the major polarizability axes $a$, $b$, and $c$, respectively. Real-valued and wavelength-independent Euler angles $\varphi$, $\theta$, and $\psi$ are used to rotate between the Cartesian laboratory frame and the Cartesian polarizability system and, together with a layer thickness $d$, complete the list of model parameters~\cite{Schubert1996,Schubertbook,Schmidt2013ellinanoscale}. Here, the complex dielectric function parameters have been calculated wavelength by wavelength, i.e., $\varepsilon_{a}$, $\varepsilon_{b}$, and $\varepsilon_{c}$ are obtained at each wavelength, independent from all other spectral data points. Hence, no physical line-shape model was implemented, which allows for an unbiased extraction of the intrinsic dielectric function tensor. Independent Kramers-Kronig consistency tests can then be done with the individual dielectric function parameters~\cite{Dressel2008}.

Data analysis requires nonlinear regression methods, where measured and calculated Mueller matrices are matched as closely as possible by varying the above-mentioned model parameters, thereby minimizing a weighted test function. Both samples discussed here have been analyzed using a multisample-configuration analysis scheme, i.e., all nine in-plane orientations have been included in the regression analysis and only the azimuth Euler angle $\varphi$ has been set according to the measurement configuration ($\varphi_{n+1}=\varphi_{n}+40^{\circ}$)~\cite{Schmidt2013ellinanoscale}.

\section{Results and Discussions}

Figure~\ref{fig:spectrum} depicts representative experimental and best-match calculated Mueller matrix spectra for both samples under investigation. Spectra are plotted for one angle of incidence and one in-plane orientation (see the figure caption for details). Both samples exhibit anisotropic optical behavior, as is evident from the off-diagonal Mueller matrix elements, and the degree of anisotropy is larger for the rhombohedral BiFeO$_3$ film. As a result of the nonlinear regression fitting, film thicknesses of $40.8\pm0.1$ and $32.7\pm0.3$~nm have been determined for BiFeO$_3$ on SrTiO$_3$ and LaAlO$_3$, respectively. The latter is in very good agreement with a thickness of 31~nm measured by x-ray diffraction. The thicknesses of the surface roughness layers are $2.5\pm0.1$ and $3.0\pm0.2$~nm, respectively.

\subsection{Dielectric function tensors}

The wavelength-by-wavelength extracted dielectric function tensor as well as a schematic representation of the major polarizability coordinate system orientation with respect to crystallographic axes are presented in Fig.~\ref{fig:eps}. In both cases a large degree of birefringence and dichroism is observed, and while BiFeO$_3$ on LaAlO$_3$ indicates a uniaxial character ($\varepsilon_a \approx \varepsilon_b \neq \varepsilon_c$), BiFeO$_3$ on SrTiO$_3$ has biaxial optical properties ($\varepsilon_a \neq \varepsilon_b \neq \varepsilon_c$).
Within the investigated spectral range BiFeO$_3$ on SrTiO$_3$ exhibits an average of 30\% and 44\% less birefringence and dichroism, respectively, compared to the highly strained BiFeO$_3$ on LaAlO$_3$. Similar to previous reports on bulklike rhombohedral BiFeO$_3$~\cite{Choi2011,Rivera1997}, a negative birefringence ($n_c<n_a,n_b$, $n_j={\rm Re}\{\sqrt{\varepsilon_j}\}$) is observed below 2.5~eV for both films. Interestingly, the birefringence is significantly more pronounced for tetragonal-like BiFeO$_3$ ($\approx70$\% at 1.5~eV), which is in contrast to first-principles DFT, where the birefringence was calculated to be nearly identical between both phases.~\cite{Dong2013}
While the overall shapes of both dielectric function tensors show similarities, the peak positions in the case of BiFeO$_3$ on LaAlO$_3$ are blueshifted with respect to BiFeO$_3$ on SrTiO$_3$. Besides that, the behavior in the low-energy range shows some interesting characteristics: On LaAlO$_3$, Im$\{\varepsilon_{a}\}$ and Im$\{\varepsilon_{b}\}$ below 1.8~eV and Im$\{\varepsilon_{c}\}$ below 1.2 eV are zero (within the experimental error), while this is not observed within the measured spectral range for the film on SrTiO$_3$.

%Please note that there is very low sensitivity to the $c$-axis.
%, with a refractive index $n = [(|{\rm Re}\{\varepsilon\}|+{\rm Im}\{\varepsilon\})/2]^{0.5}$)
%($\varepsilon_j=\varepsilon_{1j}+i\varepsilon_{2j}$)

Furthermore, optical modeling revealed that even though BiFeO$_3$ on LaAlO$_3$ exhibits uniaxial-like optical properties, the major polarizability axis $\varepsilon_{c}$ (optic axis) does not coincide with the [001] crystallographic axis but rather is tilted away from the surface normal by $\theta = 7.4\pm 0.9^{\circ}$. This tilt has been confirmed with careful PFM studies as well as measurements of in- and out-of-plane polarization components~\cite{Zhang2011,Chen2012}, resulting in a value of $\leq9^{\circ}$. The small unintentional substrate miscut results in an overall preferred domain alignment responsible for this net polarization tilt. The Euler angle $\psi=0^{\circ}$ was not included in the analysis, resulting in the major polarizability axis $a$ being in plane~\cite{footnote}.

For BiFeO$_3$ on SrTiO$_3$, the Euler angles $\theta = 42.8^{\circ}$ and $\psi = 0^{\circ}$ have not been included in the regression analysis, and the calculated angle $\varphi=41.865\pm0.007^{\circ}$ indicates that the polarization is along the [111] direction~\cite{Jang2008,footnote}.

\begin{figure}[tbp]
	\includegraphics[width=.5\textwidth, clip, trim=30 20 0 50]{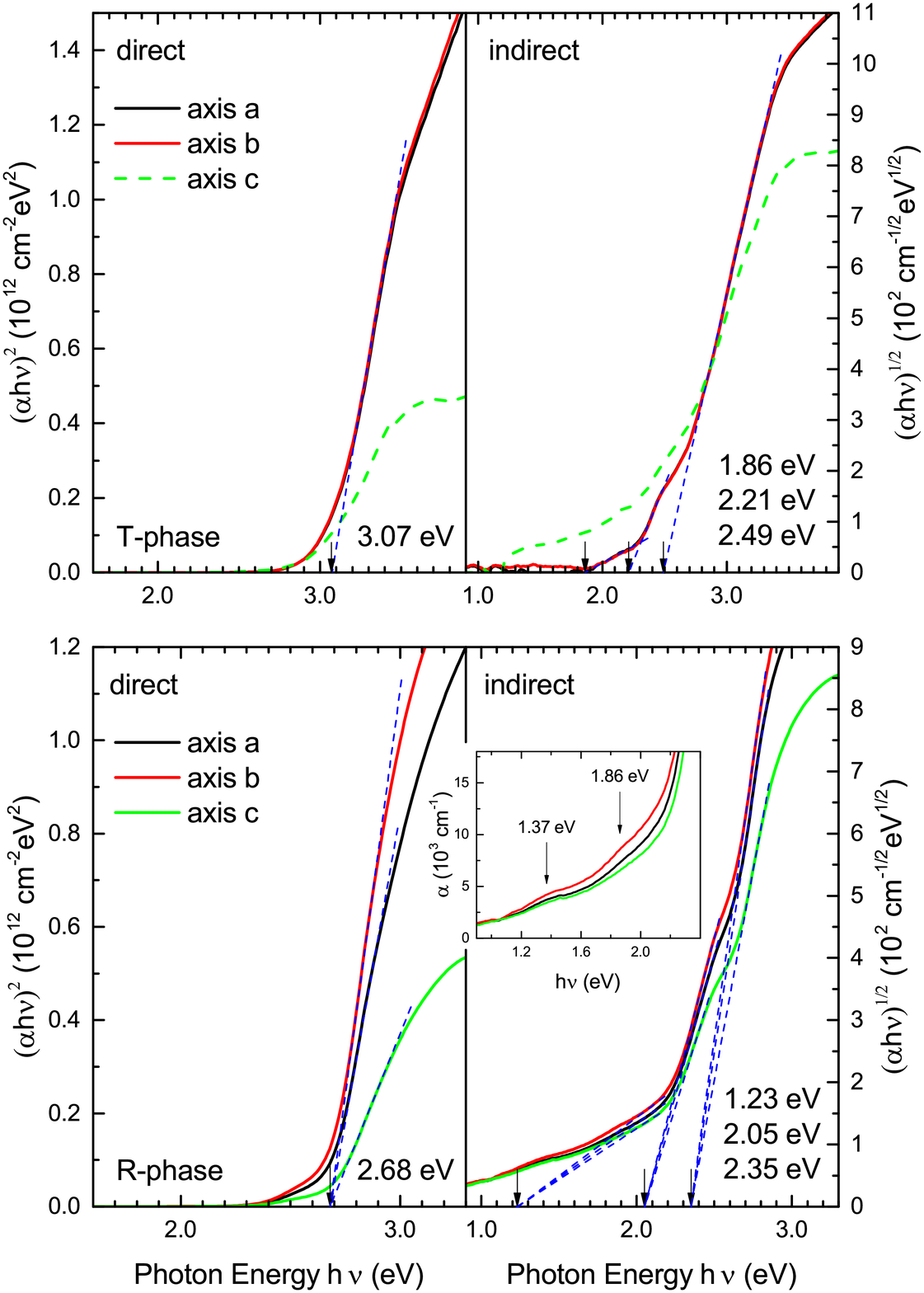}
	\caption{Plots of $(\alpha h \nu)^{2}$ and $(\alpha h \nu)^{1/2}$ vs photon energy visualizing direct and indirect transitions, respectively, for BiFeO$_3$ on LaAlO$_3$ (top) and BiFeO$_3$ on SrTiO$_3$ (bottom). The inset shows the absorption coefficient $\alpha$ vs photon energy highlighting two subtle excitations for \emph{R}-phase BiFeO$_3$. No such excitations are observed for \emph{T}-phase BiFeO$_3$.}
	\label{fig:transitions}
\end{figure}

%\begin{figure}[tbp]
%	\includegraphics[width=.5\textwidth]{2LAOdirect_indirect_allowed.eps}
%	\caption{Plots of $(\alpha h \nu)^{2}$ and $(\alpha h \nu)^{1/2}$ versus photon energy depicting direct and indirect transitions for BiFeO$_3$ on LaAlO$_3$, respectively. No sub-band gap excitations are observed below the indirect band gap.}
%	\label{fig:LAO_gap}
%\end{figure}

%\begin{figure}[tbp]
%	\includegraphics[width=.5\textwidth]{2STOdirect_indirect_allowed.eps}
%	\caption{Plots of $(\alpha h \nu)^{2}$ and $(\alpha h \nu)^{1/2}$ versus photon energy depicting direct and indirect transitions for BiFeO$_3$ on SrTiO$_3$, respectively. The inset shows the absorption coefficient $\alpha$ versus photon energy highlighting two subtle sub-band gap excitations.}
%	\label{fig:STO_gap}
%\end{figure}

\subsection{Transitions and Excitations}

The dielectric function tensor enables computation of many relevant sample properties as well as the absorption coefficient $\alpha$, which in turn allows for the determination of transition characteristics. Plots of the square and square root of $\alpha h \nu$ allow for a graphical determination of direct and indirect transitions, respectively, by extrapolating the linear regimes to $\alpha h \nu = 0$~\cite{Johnson1967}. For an indirect band gap at room temperature, two linear branches should be observable, corresponding to phonon emission and absorption~\cite{Pankove1975}.

%$\alpha=4\pi k \lambda^{-1}$
%Such plots are depicted in Figs.~\ref{fig:LAO_gap} and ~\ref{fig:STO_gap} BiFeO$_3$ on LaAlO$_3$ and BiFeO$_3$ on SrTiO$_3$, respectively.

Figure~\ref{fig:transitions} depicts plots of $(\alpha h \nu)^{2}$ and $(\alpha h \nu)^{1/2}$ versus photon energy $h \nu$. For BiFeO$_3$ on LaAlO$_3$ (top panel), the first direct transition is observed at 3.07~eV by extrapolating both axes $a$ and $b$; this energy position is in agreement with previous reports for a direct gap~\cite{Chen2010,Himcinschi2015}. Since this transition is only to Fe $3d_{xy}$ orbitals, axis $c$ is not considered for the graphical analysis~\cite{Dong2013}.
For BiFeO$_3$ on SrTiO$_3$, the first direct transition is found by extrapolation of all three axes at 2.68~eV (Fig.~\ref{fig:transitions}, bottom panel). This is in agreement with previous reports for comparable \emph{R}-phase samples~\cite{Kumar2008,Ihlefeld2008,Himcinschi2010} and corresponds to a redshift of around 390~meV with respect to BiFeO$_3$ on LaAlO$_3$.

%Note that the graph for axis $c$ is omitted due to some experimental data noise resulting from a lower sensitivity to the out-of-plane component.

In addition to the first direct transition, there are several linear regimes in the $(\alpha h \nu)^{1/2}$ plot that can be extrapolated to determine other excitations (Fig.~\ref{fig:transitions}). Particularly, two linear branches are observed, potentially representing phonon absorption (${\rm E_g-E_p} = 1.86$~eV) and emission (${\rm E_g+E_p} = 2.21$~eV) in \emph{T}-phase BiFeO$_3$. This would set the indirect band gap $\rm E_g$ at 2.035 eV with an involved phonon of 175~meV. An analogous analysis for BiFeO$_3$ on SrTiO$_3$ would result in an indirect band gap at $\rm E_g = 1.64$~eV with a phonon energy of $\rm E_p=410$~meV. However, the required optical phonon energies, particularly in the case of \emph{R}-phase BiFeO$_3$, are too high for such a scenario. Nevertheless, since the branch related to phonon absorption is usually very subtle it would be reasonable to assume that extrapolation of a single linear regime results in an indirect gap with involved phonon emission ${\rm E_g+E_p}$.~\cite{MacFarlane1955} This, however, leaves the graphical determination of a possible indirect band gap inconclusive. Any of the three determined values would be a candidate (with sub-band gap excitations where applicable), but none of the theoretical calculations have placed an indirect gap energetically so far below the first direct gap to make an assumption~\cite{Wang2009,Palai2008}.

%Hence, similar to the first direct transition, the indirect gap would be shifted by 395~meV towards lower energies compared to BiFeO$_3$ on LaAlO$_3$.
%Still, plots involving the absorption coefficient are interesting to discover spectral features.

The graphically determined transitions at 2.35 and 2.49~eV for the \emph{R} and \emph{T} phase, respectively, are responsible for small shoulders in the dielectric function and have been observed in many previous studies. It was proposed that these excitations are likely defect related and, due to the presence of a moderately shallow oxygen vacancy state, approximately 0.3-0.6~eV below the direct band gap~\cite{Hauser2008,Clark2009}. The next lower-energy excitations (2.05 and 2.21~eV for the \emph{R} and \emph{T} phase, respectively) may then be assigned to a dipole-forbidden on-center $t_{1g}(\pi)\rightarrow t_{2g}$ charge transfer transition, which has been predicted to be redshifted by about 0.8~eV with respect to the respective direct gap~\cite{Pisarev2009}.

Interestingly, for \emph{R}-phase BiFeO$_3$, two additional subtle excitations below 2~eV are identified and emphasized in the inset of Fig.~\ref{fig:transitions}. The peak positions have been determined by a dedicated nonlinear regression analysis of the wavelength-by-wavelength extracted Im$\{\varepsilon\}$ and are at 1.37 and 1.86~eV. These energy positions are in very good agreement with previously observed sub-band gap transitions as a result of transmittance measurements on a bulk-like rhombohedral single crystal. The two very weak peaks can be attributed to many-particle transition bands and specifically comprising a pure exciton transition ${}^6A_{1g} \rightarrow {}^4T_{1g}$ and another ${}^6A_{1g} \rightarrow {}^4T_{2g}$ on-site Fe$^{3+}$ crystal-field transition barely allowed by spin-orbit coupling~\cite{Xu2009,Ogawa2004,Galuza1998}. Together, these sub-band gap excitations are responsible for the absorption onset for \emph{R}-phase BiFeO$_3$ being outside the measured spectral range.

These transitions are not observed for the highly compressed in-plane axes $a$ and $b$ of \emph{T}-phase BiFeO$_3$, but rather the onset of absorption is at around 1.86 eV. For the strained out-of-plane axis, however, the absorption onset is significantly redshifted. Unfortunately, due to a lower experimental sensitivity to axis $c$ and hence some data noise, the subtle peak at around 2.1~eV cannot be characterized with necessary certainty, but it could be another many-particle excitation~\cite{Galuza1998}. A probable explanation for the difference in the absorption onset is that the charge transfer exciton self-trapping, which is partially responsible for the shallow absorption tail, is governed by lattice strain and is mostly suppressed here~\cite{Pisarev2009}.

%Moreover, the absorption coefficient of the strained $c$ axis might also be further influenced by self-trapped excitons as discussed below.

%$\prescript{6}{}A_{1g} \rightarrow \prescript{4}{}T_{1g}$ and $\prescript{6}{}A_{1g} \rightarrow \prescript{4}{}T_{2g}$ %requires mathtools package

\begin{figure}[tbp]
	\includegraphics[width=.45\textwidth, clip, trim=0 0 0 0]{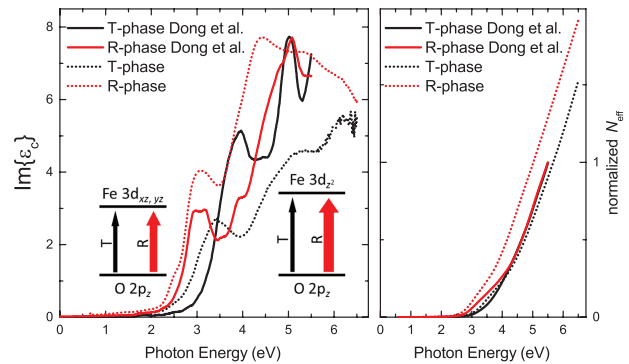}%{STO_LAO_combined2.eps}
	\caption{Imaginary parts of the dielectric function tensor $\varepsilon_c$ for \emph{R}- and \emph{T}-phase BiFeO$_3$ (left) and normalized effective carrier number $N_{\rm eff}$ (right) of this work and compared to literature values from Dong \emph{et al.}~\cite{Dong2013}. The two diagram insets show schematically the dominant excitations.}
	\label{fig:neff}
\end{figure}

%along the major polarizability axis $c$

\subsection{Comparison with theoretical calculations}

The comparison of experimentally obtained imaginary parts of the dielectric function along axis $c$ and theoretically calculated values taken from Dong \emph{et al.}~\cite{Dong2013} is shown in Fig.~\ref{fig:neff}. The dielectric function of each phase below 6~eV is dominated by two intense bands (around 3 and 4.5~eV for \emph{R} phase, and 3.5 and 5~eV for \emph{T} phase), which can be assigned to ${\rm O}~2p_z \rightarrow  {\rm Fe}~3d_{xz,yz}$ and ${\rm O}~2p_z \rightarrow  {\rm Fe}~3d_{z^2}$ electron excitations~\cite{Dong2013,Pisarev2009}. In general, it is important to note that a peak in Im$\{\varepsilon\}$ does not necessarily correspond to a single interband transition only as multiple direct or indirect transitions may be found in the band structure with an energy corresponding to or very close to the same peak.\cite{Wang2009}

Experimentally, the two main \emph{T}-phase peak positions are blueshifted with respect to the \emph{R} phase by 0.3 and 0.8~eV for low- and high-energy excitations, respectively. This blueshift can be explained by local symmetry breaking of the FeO$_6$ octahedral due to the high compressive strain.~\cite{Dong2013,Chen2010} The thin and thick arrows in the inset of Fig.~\ref{fig:neff} depict schematically the experimentally found relative transition strengths for \emph{T}- and \emph{R}-phase BiFeO$_3$. The additional \emph{T}-phase electronic feature centered at 6.2~eV may be due to a strongly hybridized majority channel ${\rm O}~2p_z + {\rm Fe}~3d_{z^2}\rightarrow {\rm Bi}~2p$ state excitation.~\cite{Dong2013,Chen2010} It is assumed that for \emph{R}-phase BiFeO$_3$ such a transition is responsible for the 0.8~eV redshifted small peak at 5.4~eV.

For \emph{R}-phase BiFeO$_3$, the theoretically determined low-energy \emph{p-d} transition strength is slightly underestimated while the high-energy one is in excellent agreement with the experiment; however, the peak position is blueshifted by approximately 0.7~eV. Note that all dielectric functions obtained by DFT had been shifted equally to match the location of the low-energy band of \emph{R}-phase BiFeO$_3$.
Regarding the energy positions of the main transitions, the situation is the opposite for \emph{T}-phase BiFeO$_3$. While the high-energy excitation peak around 5~eV matches very well with the experiment, the low-energy \emph{p-d} transition is blue-shifted by about 0.5~eV. Both calculated excitation strengths are significantly overestimated compared to the experiment.

%For T-phase BiFeO$_3$ the situation is opposite in terms of energy positions of the main transitions.

%; and $t_{2u}(\pi) \rightarrow  t_{2g}$ and $t_{1u}(\pi) \rightarrow  t_{2g}$ according to Pisarev et al.~\cite{Pisarev2009}
%from ${\rm O}~2p_z$ to ${\rm Fe}~3d_{xz,yz}$ and ${\rm Fe}~3d_{z^2}$, respectively.~\cite{Pisarev2009,Dong2013}

In addition, Fig.~\ref{fig:neff} shows the normalized effective number of carriers $N_{\rm eff}$ defined as the integration of the optical conductivity $\sigma$ over the measured frequency range~\cite{Takenaka2000}. Here, the averaged optical conductivity along all major polarizability axes is taken into account.
Results from the DFT calculations show that except for a small range between 3 and 4~eV, an identical amount of carriers for \emph{R}- and \emph{T}-phase BiFeO$_3$ are participating in optical transitions. The experimentally observed situation, however, is different and the number of effective carriers is significantly larger for \emph{R}-phase BiFeO$_3$ within the measured spectral range. This means a substantial spectral weight transfer to higher energies ($>6.5$~eV) occurs for \emph{T}-phase BiFeO$_3$. Consequently, the current first-principles DFT calculations with exchange-correlation functionals do not fully account for all many-body effects that may influence the electronic structure. Hence, in order to improve the theoretical calculations of optical spectra and thus allow for a more comprehensive interpretation of this complex multiferroic compound, many-body effects such as exchange and correlation interactions must be further refined and especially excitonic excitations taken into account.

%$\sigma(\omega)= \omega \varepsilon_0 {\rm Im}\{\varepsilon\}$

\section{Conclusions}

In conclusion, we have carefully determined and investigated the anisotropic dielectric function tensor of tetragonal-like and rhombohedral-like multiferroic BiFeO$_3$ phases. Optical modeling of spectroscopic Mueller matrix ellipsometry data enabled the quantification of strain-driven birefringence and dichroism as well as band structure changes in terms of transition energy shifts.

It is found that the birefringence in the low-energy range ($<2.5$~eV) is significantly larger for tetragonal-like BiFeO$_3$, which is expected but is against previous calculations. The first direct transition has been determined for each of the two BiFeO$_3$ phases, and they are separated by approximately 0.4~eV. Further excitations have been quantified, and the difficulty of a graphical identification of a potential indirect band gap is discussed.
The unbiased wavelength-by-wavelength analysis also allowed for a characterization of subtle sub-band gap excitations in rhombohedral BiFeO$_3$ on SrTiO$_3$, which are not present in the same manner in highly strained tetragonal BiFeO$_3$ on LaAlO$_3$.

Furthermore, optical modeling and polarization component measurements revealed that the major polarization of BiFeO$_3$ on LaAlO$_3$ within a bivariant in-plane configuration is tilted away from the crystallographic [001] axis and the surface normal.

Comparing our experimental data to recent density functional theory calculations has exposed shortcomings of the used first-principles algorithms. Highly strained tetragonal-like BiFeO$_3$ on LaAlO$_3$ exhibits a substantial spectral weight transfer to higher energies beyond 6.5~eV compared to its rhombohedral-like counterpart, which is not predicted by the current quantum mechanical modeling method. These discrepancies likely stem from many-body effects, such as exchange and correlation interactions (probably not correctly accounted for) and particularly excitonic effects (typically not regarded at all), when calculating the optical spectra of BiFeO$_3$.

The provided rigorous anisotropic optical modeling and the full dielectric function tensor for both BiFeO$_3$ phases will be highly useful to improve \emph{ab initio} calculations to better understand multiferroic materials.

%The strain-driven optical anisotropy and band-gap changes together with the fact that rhombohedral and tetragonal phases can be reversibly controlled by an external electric field~\cite{Zhang2013} may allow for development of piezoelectrically driven electro-optic devices.

\begin{acknowledgements}
The authors would like to thank P.K. Gogoi and P.E. Trevisanutto for fruitful discussions and J.C.W. Lim for technical assistance.
This work is supported by the Singapore National Research Foundation under its Competitive Research Funding (NRF-CRP 8-2011-06 and NRF2008NRF-CRP002024), MOE-AcRF Tier-2 (MOE2010-T2-2-121), NUS-YIA, and FRC.
\end{acknowledgements}

\end{document}